\documentclass[universe,article,accept,moreauthors,pdftex,12pt,a4paper]{mdpi}
\lastpage{\pageref*{LastPage}}
\doinum{10.3390/universe1030412}
\pubvolume{1}
\pubyear{2015}
\externaleditor{Academic Editor: Gonzalo J. Olmo}
\history{Received: 27 October 2015 / Accepted: 9 November 2015  / Published: 16 November 2015}
\pdfoutput=1

\usepackage{soul}

\def\d{\mathrm{d}}

\Title{Thermodynamic Analysis of Non-Linear Reissner-Nordstr\"{o}m Black Holes}

\Author{Jose A.~R.~Cembranos $^{1,\dagger}$, \'Alvaro de la Cruz-Dombriz $^{2,\dagger,}$* and Javier Jarillo $^{3,\dagger}$}

\address{%
$^{1}$ Departamento de F\'{\i}sica Te\'orica I, Universidad Complutense de Madrid, E-28040 Madrid, Spain; {E-Mail: cembra@fis.ucm.es} \\
$^{2}$ Astrophysics, Cosmology and Gravity Centre (ACGC), Department of Mathematics and Applied Mathematics, University of Cape Town, Rondebosch 7701, Cape Town, South Africa \\
$^{3}$ Departamento de F\'{\i}sica At{\'o}mica, Molecular y Nuclear, Universidad Complutense de Madrid, E-28040 Madrid, Spain; {E-Mail: jjarillo@ucm.es}}

\contributed{$^\dagger$ {These authors contributed equally to this work.}}

\corres {E-Mail: alvaro.delacruzdombriz@uct.ac.za.  }
\abstract{ In the present article we study the Inverse Electrodynamics Model. \mbox{This model} is a gauge and parity invariant non-linear Electrodynamics theory, which respects the conformal invariance of standard Electrodynamics. This modified Electrodynamics model, when minimally coupled to General Relativity, is compatible with static and spherically symmetric Reissner-Nordstr\"{o}m-like black-hole solutions. However, these black-hole solutions present more complex thermodynamic properties than their Reissner-Nordstr\"{o}m black-hole solutions counterparts in standard Electrodynamics. In particular, in the Inverse Model a new stability region, with both the heat capacity and the free energy negative, arises. Moreover, unlike the scenario in standard Electrodynamics,
a sole transition phase is possible for a suitable choice in the set of parameters of these solutions.
}

\keyword{modified Electrodynamics; Reissner-Nordstr\"{o}m black holes; stability and thermodynamics of black holes}

\PACS{11.10.Lm; 04.40.-b; 04.70.Bw; 03.50.De}
\begin{document}

%%%%%%%%%%%%%%%%%%%%%%%%%%%%%%%%%%%%%%%%%%

\newpage

\section{Introduction: The Inverse Electrodynamics Model}

It is a well known fact that the self energy of a point charge (as, in a first approximation, an electron could be considered) in the standard Electrodynamics theory diverges. This energetic divergence has lead to the development of new non-linear Electrodynamics models, as for example the Born-Infeld \cite{born1934foundations} and Euler-Heisenberg \cite{heisenbergSet} models. The first kind among these models was first proposed by Born and Infeld in 1934  as an effective theory which behaves asymptotically as the standard Electrodynamics, but which does not suffer the aforementioned divergence of the point charge self energy. Years later, in 1936 Heisenberg and Euler, while studying light-light scattering in the context of Quantum Electrodynamics~\cite{dunne2012heisenberg, dobado2012effective}, obtained their model which also exhibits the correct asymptotic behaviour and avoids divergence at the origin. More recently, some examples of these non-linear theories have attracted a renewed attraction in the context of string theories \cite{fradkin1985non, tseytlin1997non, brecher1998bps}, and in the context of its coupling to General Relativity (GR) in the study of black-hole (BH) solutions  \cite{gibbons1995electric, ayon1998regular, fernando2003letter, fernando2006thermodynamics, hassaine2007higher, hassaine2008higher, diaz2010electrostatic, diaz2010asymptotically, banerjee2012critical-a, banerjee2012critical-b, gunasekaran2012extended, allahverdizadeh2013extremal, diaz2013thermodynamic, ruffini2013einstein, zou2014critical}.

In this investigation we shall propose the U(1) Inverse Electrodynamics (IE) Model \cite{cembranos2015reissner}, a non-linear Electrodynamics model whose action  reads (unless otherwise specified, all the magnitudes shown in this work have been expressed in Planck units, with $G = c = k_B = 4 \pi \varepsilon_0 = \hbar = 1$; furthermore, for the metric tensor we have chosen the signature $+ - - -$)
\begin{equation} \label{eq:IED_action}
S_{\rm IE} \equiv  - \int {\rm d}^4 x\sqrt{\lvert g \lvert} \mathcal{L}(X,Y) \equiv - \frac{1}{8\pi} \int \sqrt{\lvert g \lvert }\, X \left[  1 - \varepsilon \left( \frac{Y}{X} \right)^2 \right] \, \d ^4x \,,
\end{equation}
where $g$ is the determinant of the metric $g_{\mu \nu}$, and $X$ and $Y$ have been defined as the Maxwell invariants
\begin{equation} \label{eq:MaxwellInvariants}
X \equiv - \frac{1}{2} F_{\alpha \beta} F^{\alpha \beta} \,\,\,\, \text{ and } \,\,\,\, Y \equiv - \frac{1}{2} F_{\alpha \beta} \mathcal{F}^{\alpha \beta}\,,
\end{equation}
with $F_{\alpha \beta} \equiv A_{\beta;\alpha}-A_{\alpha;\beta}$ the usual electromagnetic tensor (where we employ the notation $X_{;\alpha} \equiv \nabla_\alpha X$, with $\nabla$ the standard covariant derivative),  and $\mathcal{F}_{\alpha \beta}$ its dual, $\mathcal{F}_{\alpha \beta} \equiv \frac{1}{2} \sqrt{\lvert g \lvert} \epsilon_{\alpha \beta \gamma \delta} F^{\gamma \delta}$, being $\epsilon_{\alpha \beta \gamma \delta}$ the Levi-Civita symbol. As we can see, the proposed IE Model is parity and gauge invariant (both $X$ and $Y$ are gauge invariant, and both $X$ and $Y^2$ are even). Moreover, in the limit $\varepsilon \rightarrow 0$ the standard Electrodynamics action is recovered.

\section{Reissner-Nordstr\"om-Like Black-Hole Solutions}

After coupling the IE Model to GR, the total action $S_{T}$ reads
\begin{equation} \label{eq:TotalAction}
S_{T} = S_{\rm GR} + S_{\rm IE} \,,
\end{equation}
where $S_{\rm GR}$ stands for the Einstein-Hilbert action with a cosmological constant,
\[
S_{\rm GR} \equiv \frac{1}{16\pi} \int \sqrt{\lvert g \lvert} \left( R-2 \Lambda \right) \d ^4 x\,,
\]
with $R$ and $\Lambda$ the scalar of curvature and a cosmological constant respectively.
In the following, we shall focus on the thermodynamic analysis of static and spherically symmetric solutions. Hence, we can consider as an ansatz for the metric tensor the most general static and spherically symmetric metric,
\begin{equation} \label{eq:SymmetricMetric}
\d s^2 = \lambda(r)\d t^2 - \frac{1}{\mu(r)} \d r^2 - r^2 \d \theta^2 - r^2 \sin^2 \theta \d \phi^2\,,
\end{equation}
where the functions $\lambda(r)$ and $\mu(r)$ solely depend upon the radial coordinate $r$; and for the electromagnetic tensor $F_{\mu \nu}$, we can consider an ansatz for which the unique non-zero components are
\begin{equation}
F_{tr} = - F_{rt} = E(r) \,\,\,\, \text{ and } \,\,\,\, F_{\theta \phi} = -F_{\phi \theta} =- B(r) r^2 \sin \theta\,.
\end{equation}
This ansatz coincides, in Minkowski spacetime, to the electromagnetic tensor corresponding to radial electric and magnetic fields $E(r)$ and $B(r)$ \cite{jackson1998classical}.

Performing variations of the total action Equation~\eqref{eq:TotalAction} with respect to the metric tensor, we obtain the Einstein's~equations
\begin{equation} \label{eq:EinsteinEq}
R_{\alpha \beta} - \frac{1}{2} \left( R-2\Lambda \right) g_{\alpha \beta} = 8\pi T_{\alpha \beta}\,,
\end{equation}
where $R_{\alpha \beta}$ stands for the Ricci tensor, and $T_{\alpha \beta}$ for the energy-momentum tensor,
\begin{equation} \label{eq:EnergyMomentumTensor}
T_{\alpha \beta} \equiv - \frac{2}{\sqrt{\lvert g \lvert}} \frac{\delta S_{\rm IE}}{\delta g^{\alpha \beta}}\,.
\end{equation}

On the other hand, performing variations of the total action with respect to the electromagnetic field, we obtain the Maxwell's equations in the IE Model,
\begin{equation} \label{eq:MaxwellEq}
\left( \mathcal{L}_X F^{\alpha \beta} + \mathcal{L}_Y \mathcal{F}^{\alpha \beta}\right)_{;\alpha} = 0 \,\,\,\, \text{ and } \,\,\,\, \mathcal{F}^{\alpha \beta}_{\,\,\,\,\,\,\,;\alpha}=0 \,,
\end{equation}
where we have defined $\mathcal{L}_X \equiv \frac{\partial \mathcal{L}}{\partial X}$ and $\mathcal{L}_Y \equiv \frac{\partial \mathcal{L}}{\partial Y}$. It is easy to see that in standard Electrodynamics these Maxwell's equations are reduced to the well known equations $F^{\alpha \beta}_{\,\,\,\,\,\,\,;\alpha}=0$ and $\mathcal{F}^{\alpha \beta}_{\,\,\,\,\,\,\,;\alpha}=0$.

Solving Equations~\eqref{eq:EinsteinEq} and \eqref{eq:MaxwellEq}, we finally obtain that the electric and magnetic fields read
\begin{equation}
E(r) = \frac{Q_e}{r^2} \,\,\,\, \text{ and } \,\,\,\, B(r) = \frac{Q_m}{r^2} \,,
\end{equation}
respectively, with $Q_e$ and $Q_m$ the electric and magnetic charges; and that the metric tensor corresponding to this Reissner-Nordstr\"om-like (RN-like) BH solution is
\begin{equation} \label{eq:lambda}
\lambda(r) = \mu(r) = 1 - \frac{2M}{r} + \frac{\mathcal{Q}^2}{r^2}+ \frac{1}{3} \Lambda r^2 \,,
\end{equation}
where $M$ is an integration constant, that as we will later show corresponds to the BH mass, and $\mathcal{Q}^2$ has been defined as
\begin{equation}
\mathcal{Q}^2 \equiv \left[ 1 + 4 \varepsilon \left( \frac{Q_e Q_m}{Q_e^2 - Q_m^2} \right)^2  \right] \left(Q_e^2 + Q_m^2\right) \,.
\end{equation}

In the standard Electrodynamics theory, $\varepsilon=0$ yielding $\mathcal{Q}^2$ to be just the sum of squares of the charges, and therefore
positive. However, in the IE Model this parameter $\mathcal{Q}^2$ might be negative for some values $\varepsilon <0$. Anyhow, the RN-like BH solution in the IE model contains, as in standard Electrodynamics, a~singularity at the origin.

We can now use the obtained metric, defined by Equations~\eqref{eq:SymmetricMetric} and \eqref{eq:lambda}, to get the horizon radius $r_H$ of the BH solutions, just by %obtaining the roots of $\lambda(r)$,
solving the equation of $\lambda(r_H)=0$. Doing so, and imposing that the solution must have at least one horizon, one concludes the relation
\begin{equation} \label{eq:ConditionBH}
\mathcal{Q}^2 \le r_H^2 + \Lambda r_H^4
\end{equation}
must be satisfied, where the value of $r_H$ depends on $M$, $\mathcal{Q}^2$ and $\Lambda$. Should this relation not be not satisfied, it would mean that such a set of parameters does not correspond to a BH solution, but to a naked singularity. For the $\mathcal{Q}^2 >0$ scenario, this relation tells us about the existence of an extremal BH, \emph{i.e.}, a BH solution whose outer and inner horizons merge in a unique one, with a minimum horizon radius for those electric and magnetic charges, and for which smaller RN-like BH solutions do not exist. On the other hand, provided $\mathcal{Q}^2 < 0$, no extremal BHs would exist.

Additionally, let us stress at this stage that plugging Equations~\eqref{eq:IED_action}  and \eqref{eq:lambda} in Equation~\eqref{eq:EnergyMomentumTensor}, it~is straightforward to realise that the trace of the energy-momentum tensor is zero \cite{cembranos2015reissner}. Hence, the IE Model is not just parity and gauge invariant, but it also preserves the conformal invariance of standard~Electrodynamics.

\section{Thermodynamics Analysis}

As shall be shown below, RN-like BH solutions in the IE Model present more complex thermodynamics properties than the solutions in standard Electrodynamics. Throughout this section we shall restrict ourselves to an anti-de Sitter (AdS) space, \emph{i.e.}, $\Lambda > 0$, in order to avoid normalisation problems of the Killing vectors \cite{gibbons1977cosmological}, and we shall follow the discussion presented in Ref.~\cite{cembranos2015reissner}.

First, the temperature of the BH can be computed through the relation \cite{hawking1975particle}
\begin{equation}
T = \frac{\kappa}{4\pi}\,,
\end{equation}
where $\kappa$ stands for the surface gravity of the BH,
\begin{equation}
\kappa = \lim_{r\rightarrow r_H}  \frac{\partial_r g_{tt}}{\lvert g_{tt} g_{rr} \lvert } \,,
\end{equation}
with $r_H$ the horizon radius. Replacing Equation~\eqref{eq:lambda} in this expression, we can rewrite the temperature of the BH as
\begin{equation} \label{eq:temperature}
T = \frac{1-\frac{\mathcal{Q}^2}{r_H^2}+\Lambda r_H^2}{4\pi r_H} \,.
\end{equation}

At this stage, let us mention that the conditions which guarantee a BH solution with positive temperature coincide with the relation expressed in Equation~\eqref{eq:ConditionBH}; \emph{i.e.}, the conditions for having a BH solution with positive temperature coincide with the conditions necessary to ensure a proper BH solution. Moreover, one can see the temperature of an extremal BH (provided it exists) is zero.

Once the temperature of the RN-like BH solution has been obtained, we can compute the different stability phases of the BH, defined in terms of the signs of the Helmholtz free energy and the heat capacity. For doing so, we employ the so-called Euclidean action method \cite{hartle1976path, gibbons1977action, hawking1978quantum, gibbons1978black, gibbons1993euclidean}. In this method  the time coordinate $t$ is replaced by an Euclidean time $\tau = - i t$, being  $i$ the imaginary unit. With this change of coordinates, the metric becomes Euclidean. Hence, the difference between the Euclidean action of our BH solution in an AdS space, and the Euclidean action of an empty AdS space, can be evaluated through the expression
\begin{equation} \label{eq:EuclideanActionIncomplete}
\widehat{\Delta S}_E = - \frac{1}{16 \pi} \int_\mathcal{Y} \d^4x\,\sqrt{g} \left(R - 2\Lambda - 16 \pi \mathcal{L}(X,Y) \right) \,,
\end{equation}
being $\mathcal{Y}$ the integration region. However, just following that procedure the Euclidean action involved would not be the correct one. Performing variations of Equation~\eqref{eq:EuclideanActionIncomplete} with respect to $g_{\alpha \beta}$ and $A_\alpha$, we~get
\begin{eqnarray}
\delta \widehat{\Delta S}_E & = & - \frac{1}{16 \pi} \int_\mathcal{Y} \d^4 x\, \sqrt{g} \left(R_{\alpha \beta} - \frac{1}{2} R g_{\alpha \beta} + \Lambda g_{\alpha \beta} - 8 \pi T_{\alpha \beta} \right) \delta g^{\alpha \beta} \nonumber \\
& & + 2 \int_\mathcal{Y} \d^4 x\, \sqrt{g} \left(\mathcal{L}_X F^{\alpha \beta} + \mathcal{L}_Y \mathcal{F}^{\alpha \beta} \right)_{;\alpha} \delta A_\beta \nonumber \\
& & - 2 \int_{\partial \mathcal{Y}} \d^3 x \, \sqrt{h} \, n_\alpha \left( \mathcal{L}_X F^{\alpha \beta} + \mathcal{L}_Y \mathcal{F}^{\alpha \beta} \right) \delta A_\beta \,,
\label{eq:VariationsOfEAction}
\end{eqnarray}
where $h_{\alpha \beta}$ represents the induced metric on the boundary surface $\partial \mathcal{Y}$, $h$ its determinant and $n_\alpha$ the normal vector to the boundary. The first and second terms of this Equation~\eqref{eq:VariationsOfEAction} are identically zero due to the Einstein's and Maxwell's equations, Equations~\eqref{eq:EinsteinEq} and \eqref{eq:MaxwellEq} respectively. However, the third term is zero only if $\delta A_\beta$ is zero on the boundary surface. \emph{i.e.}, Equation~\eqref{eq:EuclideanActionIncomplete} would correspond to an ensemble with fixed electric potential instead of to an ensemble of fixed electric charge \cite{caldarelli2000thermodynamics}. In order to circumvent this issue and perform the calculation in the fixed charge ensemble,
a surface term to the action must be added, which cancels the third term of Equation~\eqref{eq:VariationsOfEAction}. After the corresponding computations, the difference of Euclidean action becomes
\begin{equation} \label{eq:EuclideanAction}
\Delta S_E = \beta \frac{r_H}{4} \left( 1 +  3 \frac{\mathcal{Q}^2}{r_H^2} - \frac{1}{3} \Lambda r_H^2  \right) \,,
\end{equation}
where $\beta$ stands for the inverse of the temperature Equation~\eqref{eq:temperature}.

From this Euclidean action one can define the massive energy of the solution, $\mathcal{M}$, as the derivative of this action with respect to the inverse of the temperature $\beta$. This massive energy coincides with the integration constant $M$ of Equation~\eqref{eq:lambda}; hence, it is obvious that we can interpret the constant $M$ as the {\it mass} of the BH. On the other hand, the Helmholtz free energy is defined as the quotient between the Euclidean action Equation~\eqref{eq:EuclideanAction} and $\beta$, yielding
\begin{equation}\label{eq:FreeEnergy}
F = \frac{\Delta S_E}{\beta} = \frac{r_H}{4} \left( 1 +  3 \frac{\mathcal{Q}^2}{r_H^2} - \frac{1}{3} \Lambda r_H^2  \right) \,.
\end{equation}

Once these two energies have been obtained,  the entropy of the BH, defined as the difference $S=\beta \mathcal{M} - \beta F$, can be computed, providing $S=\pi r_H^2$, \emph{i.e.}, a quarter of the horizon area. Thus, the obtained entropy in the IE Model coincides with the Bekenstein-Hawking entropy formula of standard Electrodynamics~\cite{bekenstein1973black, hawking1975particle}, what is an expected result due to the fact that the BH's entropy is a Noether charge~\cite{wald1993black}.

Furthermore, from the entropy $S$ and the temperature $T$ of the BH it is possible to compute its heat capacity, defined as $C=T \frac{\partial S}{\partial T}$. The result is
\begin{equation} \label{eq:HeatCapacity}
C = - 2 \pi r_H^2 \frac{1 - \frac{\mathcal{Q}^2}{r_H^2} + \Lambda r_H^2}{1 - 3 \frac{\mathcal{Q}^2}{r_H^2} - \Lambda r_H^2} \,.
\end{equation}

\subsection{Stability Phases of the RN-Like Solutions}

\begin{figure}[H]
	\centering
		\includegraphics[width=0.40\textwidth]{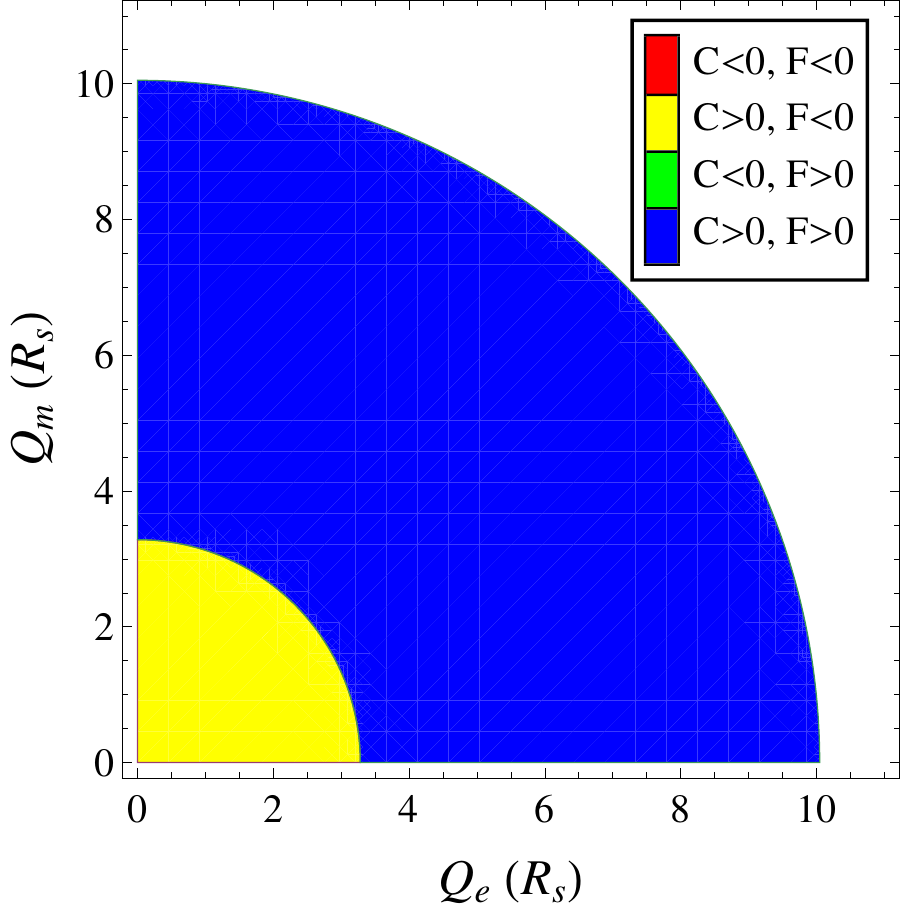}\,\,\,
		\includegraphics[width=0.40\textwidth]{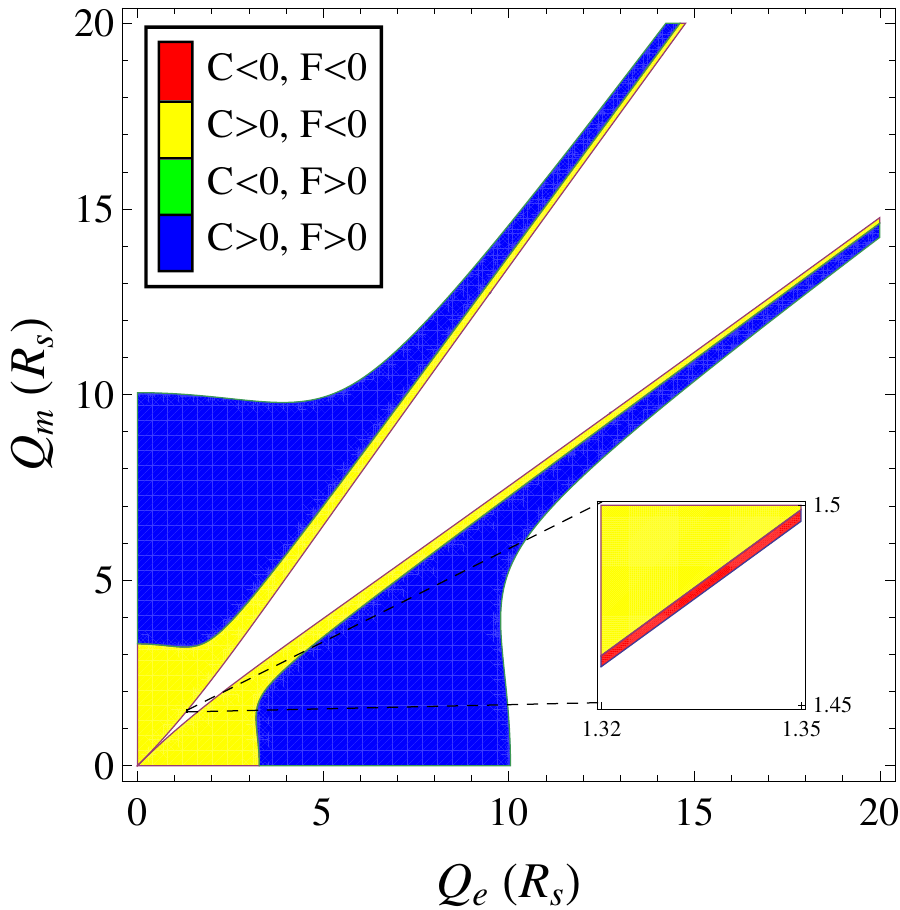}
		\caption{
		(\textbf{Left panel}) Phase stability regions for a  Reissner-Nordstr\"om (RN) black-hole (BH) solution with $r_H = 1$ $R_S$ (being $R_S$ the Schwarzschild radius of a BH with a solar mass, $R_s \sim 10^{38}$ $L_p$, in Planck units $L_p$), and $\Lambda = 100$ $R_s^{-2}$, in standard Electrodynamics;
		(\textbf{Right panel}) Phase stability regions for a RN-like BH solution with $r_H = 1$ $R_S$, $\varepsilon = -0.1$ and $\Lambda = 100$ $R_s^{-2}$ in the Inverse Electrodynamics (IE) Model. We can observe that a new phase exists, with both $F$ and $C$ negative, which does not appear in standard Electrodynamics.\vspace{-6pt}
}
	\label{fig:PhaseDiagrams}
\end{figure}

The stability regions of the BH solution can be defined in terms of the signs of the Helmholtz free energy Equation~\eqref{eq:FreeEnergy} and the heat capacity Equation~\eqref{eq:HeatCapacity} \cite{hawking1983thermodynamics}. On the one hand, BHs with $F>0$ will tend to decay to pure radiation via tunneling, contrarily to BHs with $F<0$. On the other hand, BHs with $C<0$ are unstable under acquiring mass, whereas BHs with $C>0$ are stable. In the left panel of Figure~\ref{fig:PhaseDiagrams} we have represented the phase diagram of a RN BH solution in standard Electrodynamics, whereas in the right panel it is represented the phase diagram of a RN-like BH solution in the IE Model for negative $\varepsilon$.
From this figure, we can observe the IE Model presents new features, namely,
that a new phase  with both $C$ and $F$ negative arises in the IE Model. This phase is actually absent  in standard Electrodynamics. There is also another phase that is not depicted in the figure, corresponding to negative heat capacity and positive free energy, but it appears in both models for lower values of $\Lambda$ (and, particularly in flat space, $\Lambda = 0$) \cite{cembranos2015reissner}.

\subsection{Classification of RN-Like BH Solutions in Terms of the Number of Phase Transitions}

\begin{figure}[H]
  \centering
      \includegraphics[width=0.6\textwidth]{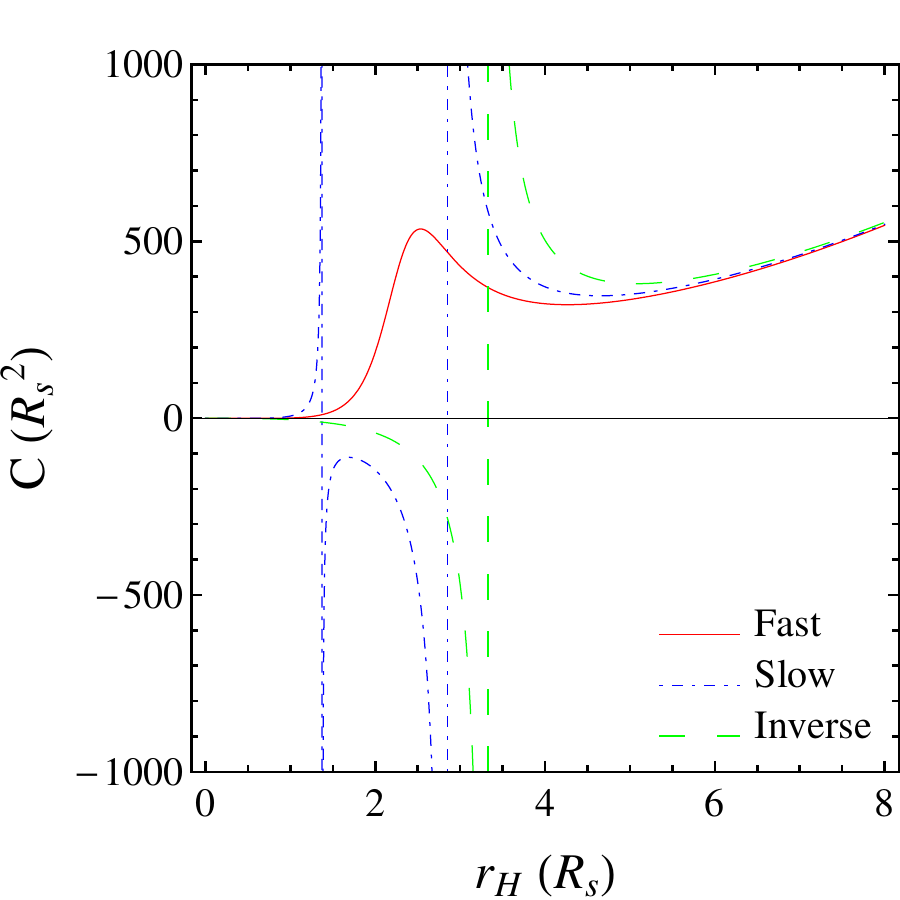}
  \caption{Heat capacity as function of the horizon radius for different classes of RN-like BH solutions. In the IE Model we can distinguish between \emph{fast} BHs, with no phase transitions; \emph{slow} BHs, with two phase transitions; and \emph{inverse} BHs, with a unique phase transition. The parameters used for each BH solution were: (1) {\it Fast} (red solid line), $\varepsilon = -0.1$, $Q_e = 1$~$R_s$, $Q_m=0.1$~$R_s$, $\Lambda = 0.1$ $R_s^{-2}$; (2) {\it Slow} (blue dot-dash line), $\varepsilon = -0.1$, $Q_e = 0.7$ $R_s$, $Q_m=0.2$ $R_s$, $\Lambda = 0.1$ $R_s^{-2}$; (3) {\it Inverse} (green dashed line), $\varepsilon = -0.1$, $Q_e = 0.5$ $R_s$, $Q_m=0.4$ $R_s$, $\Lambda = 0.1$ $R_s^{-2}$. \label{fig:PhaseTransitions}\vspace{-6pt}}
\end{figure}

Additionally, it is possible to provide a classification in terms of the number of phase transitions
 that the BH solutions host. In other words, the number of divergences of the heat capacity Equation~\eqref{eq:HeatCapacity} as a function of the horizon radius \cite{jardim2012thermodynamics, banerjee2012critical-a, banerjee2012critical-b, niu2012critical, cembranos2014kerr}. Thus, in the IE Model  the following nomenclature can be used  \cite{cembranos2015reissner}: \emph{fast} BHs, with no phase transitions, corresponding to $\mathcal{Q}^2>1/12$ $\Lambda$; \emph{slow} BHs, with two phase transitions, corresponding to $0 < \mathcal{Q}^2 < 1/12$ $\Lambda $; and \emph{inverse} BHs, with a unique phase transition, corresponding to $\mathcal{Q}^2<0$. The latter class of BH solutions does not appear in standard Electrodynamics, but it also arises in other non-linear theories, as for instance in the Born-Infeld model \cite{fernando2006thermodynamics}. In Figure~\ref{fig:PhaseTransitions} we have provided one example of each of these classes of BHs. \vspace{-6pt}

\section{Conclusions}

In this paper we have studied the Inverse Electrodynamics Model: a non-linear Electrodynamics model that respects the parity, gauge and conformal invariances of standard Electrodynamics, and which, when coupled to General Relativity, also provides static and spherically symmetric (Reissner-Nordstr\"{o}m-like) black-hole solutions.

Furthermore, using the Euclidean action method, we have performed a thermodynamic analysis of these Reissner-Nordstr\"{o}m-like solutions, studying both the stability regions of the solutions and the number of phase transitions. First, we have observed that the Inverse Model supports a new stability region of black-hole solutions, with both the heat capacity and the free energy negative. This kind of solutions will be then less energetic than pure radiation and consequently they do not decay via tunneling. Moreover, since they have negative heat capacities, they will decrease their temperature under acquiring mass. Finally, we have also seen that in the Inverse Electrodynamics Model black holes with a sole phase transition exist, a non-existing feature in standard Electrodynamics. Hence, we conclude that although the gravitational solutions of both Electrodynamics models
could be thought to be the same, the thermodynamics properties greatly differ. Further investigation and generalisation of this model are in progress.

%%%%%%%%%%%%%%%%%%%%%%%%%%%%%%%%%%%%%%%%%%

%\section{Experimental Section}
%%% Only for the journal Gels: Please place the Experimental Section after the Conclusions
%
%Main text paragraph.
%
%Main text paragraph.
%
%%%%%%%%%%%%%%%%%%%%%%%%%%%%%%%%%%%%%%%%%%%
%
%\subsection{This is a Subsection Heading}
%
%Main text paragraph.
%
%Main text paragraph.
%
%
%\subsubsection{This is a Subsubsection Heading}
%
%Main text paragraph.

%%%%%%%%%%%%%%%%%%%%%%%%%%%%%%%%%%%%%%%%%%

%%%%%%%%%%%%%%%%%%%%%%%%%%%%%%%%%%%%%%%%%%

\acknowledgments{Acknowledgments}

Jose A. R. Cembranos and Alvaro De La Cruz-Dombriz acknowledge financial support from MINECO (Spain) projects  FIS2014-52837-P, FPA2014-53375-C2-1-P and Consolider-Ingenio MULTIDARK CSD2009-00064. Alvaro De La Cruz-Dombriz also acknowledge support  from the University of Cape Town (UCT) Launching Grants programme.
Javier Jarillo acknowledges the funding support of the Spanish MECD FPU grant FPU13/02934.

%%%%%%%%%%%%%%%%%%%%%%%%%%%%%%%%%%%%%%%%%%

\authorcontributions{Author Contributions}

All authors have equally contributed to this paper, and have read and approved the final version.
%Required if more than one author. Authorship must include and be strictly limited to researchers who have substantially contributed to the reported work. Please carefully review our criteria regarding the Qualification for Authorship: \web /instructions.

%%%%%%%%%%%%%%%%%%%%%%%%%%%%%%%%%%%%%%%%%%

\conflictofinterests{Conflicts of Interest}

The authors declare no conflict of interest.

%%=================================================================
%% References: Variant A
%%=================================================================
%% Back Matter (References and Notes)
%%----------------------------------------------------------
%% Style and layout of the references
%\bibliographystyle{mdpi}
%\makeatletter
%\renewcommand\@biblabel[1]{#1. }
%\makeatother
%
%\begin{thebibliography}{999} % if there are less than 10 entries, enter a one digit number
%
%% Reference 1
%\bibitem{ref-journal}
%Lastname, F.; Author, T. The title of the cited article. {\em Journal Abbreviation} {\bf 2008}, {\em 10}, 142-149.
%
%% Reference 2
%\bibitem{ref-book}
%Lastname, F.F.; Author, T. The title of the cited contribution. In {\em The Book Title}; Editor, F., Meditor, A., Eds.; Publishing House: City, Country, 2007; pp. 32-58.
%
%\end{thebibliography}

%=================================================================
% References:  Variant B
%=================================================================
 %Use the following option to include external BibTeX files:
%\bibliography{lite}
%\bibliography{biblio_universe-v02}

\begin{thebibliography}{----}
\providecommand{\natexlab}[1]{#1}

\bibitem[Born and Infeld(1934)]{born1934foundations}
Born, M.; Infeld, L.
\newblock Foundations of the new field theory.
\newblock {\em Proc. R. Soc. Lond. Ser. A Contain.
  Pap. Math. Phys. Character} {\bf 1934}, \emph{144}, 425--451.

\bibitem[Heisenberg and Euler(1936)]{heisenbergSet}
Heisenberg, W.; Euler, H.
\newblock Consequences of Dirac's theory of positrons.
\newblock {\em Z. Phys.} {\bf 1936}, {\em 98},~714--732.

\bibitem[Dunne(2012)]{dunne2012heisenberg}
Dunne, G.V.
\newblock The Heisenberg-Euler effective action: 75 years on.
\newblock {\em Int. J. Mod. Phys. A} {\bf 2012}, {\em
  27},~1260004.

\bibitem[Dobado \em{et~al.}(2012)Dobado, G{\'o}mez-Nicola, Maroto, and
  Pelaez]{dobado2012effective}
Dobado, A.; G{\'o}mez-Nicola, A.; Maroto, A.L.; Pelaez, J.R.
\newblock {\em Effective Lagrangians for the Standard Model}; Springer Science
  \& Business Media: {Berlin, Germany},  2012.

\bibitem[Fradkin and Tseytlin(1985)]{fradkin1985non}
Fradkin, E.; Tseytlin, A.A.
\newblock Non-linear electrodynamics from quantized strings.
\newblock {\em Phys. Lett. B} {\bf 1985}, {\em 163},~123--130.

\bibitem[Tseytlin(1997)]{tseytlin1997non}
Tseytlin, A.A.
\newblock On non-abelian generalisation of the Born-Infeld action in string
  theory.
\newblock {\em Nucl.~Phys.~B} {\bf 1997}, {\em 501},~41--52.

\bibitem[Brecher(1998)]{brecher1998bps}
Brecher, D.
\newblock BPS states of the non-Abelian Born-Infeld action.
\newblock {\em Phys. Lett. B} {\bf 1998}, {\em 442},~117--124.

\bibitem[Gibbons and Rasheed(1995)]{gibbons1995electric}
Gibbons, G.; Rasheed, D.
\newblock Electric-magnetic duality rotations in non-linear electrodynamics.
\newblock {\em Nucl.~Phys. B} {\bf 1995}, {\em 454},~185--206.

\bibitem[Ay\'on-Beato and Garc\'{\i}a(1998)]{ayon1998regular}
Ay\'on-Beato, E.; Garc\'{\i}a, A.
\newblock Regular Black Hole in General Relativity Coupled to Nonlinear
  Electrodynamics.
\newblock {\em Phys. Rev. Lett.} {\bf 1998}, {\em 80},~5056--5059.

\bibitem[Fernando and Krug(2003)]{fernando2003letter}
Fernando, S.; Krug, D.
\newblock Letter: Charged Black Hole Solutions in Einstein-Born-Infeld Gravity
  with a Cosmological Constant.
\newblock {\em Gen. Relativ. Gravitat.} {\bf 2003}, {\em
  35},~129--137.

\bibitem[Fernando(2006)]{fernando2006thermodynamics}
Fernando, S.
\newblock Thermodynamics of Born-Infeld\char21{}anti-de Sitter black holes in
  the grand canonical ensemble.
\newblock {\em Phys. Rev. D} {\bf 2006}, {\em 74},~104032.

\bibitem[Hassa{\"\i}ne and Mart{\'\i}nez(2007)]{hassaine2007higher}
Hassa{\"\i}ne, M.; Mart{\'\i}nez, C.
\newblock Higher-dimensional black holes with a conformally invariant Maxwell
  source.
\newblock {\em Phys. Rev. D} {\bf 2007}, {\em 75},~027502.

\bibitem[Hassa{\"\i}ne and Mart{\'\i}nez(2008)]{hassaine2008higher}
Hassa{\"\i}ne, M.; Mart{\'\i}nez, C.
\newblock Higher-dimensional charged black hole solutions with a nonlinear
  electrodynamics source.
\newblock {\em Class. Quantum Gravity} {\bf 2008}, {\em 25},~195023.

\bibitem[Diaz-Alonso and
  Rubiera-Garcia(2010{\natexlab{a}})]{diaz2010electrostatic}
Diaz-Alonso, J.; Rubiera-Garcia, D.
\newblock Electrostatic spherically symmetric configurations in gravitating
  nonlinear electrodynamics.
\newblock {\em Phys. Rev. D} {\bf 2010}, {\em 81},~064021.

\bibitem[Diaz-Alonso and
  Rubiera-Garcia(2010{\natexlab{b}})]{diaz2010asymptotically}
Diaz-Alonso, J.; Rubiera-Garcia, D.
\newblock Asymptotically anomalous black hole configurations in gravitating
  nonlinear electrodynamics.
\newblock {\em Phys. Rev. D} {\bf 2010}, {\em 82},~085024.

\bibitem[Banerjee and Roychowdhury(2012{\natexlab{a}})]{banerjee2012critical-a}
Banerjee, R.; Roychowdhury, D.
\newblock Critical phenomena in Born-Infeld AdS black holes.
\newblock {\em Phys.~Rev.~D} {\bf 2012}, {\em 85},~044040.

\bibitem[Banerjee and Roychowdhury(2012{\natexlab{b}})]{banerjee2012critical-b}
Banerjee, R.; Roychowdhury, D.
\newblock Critical behavior of Born-Infeld AdS black holes in higher
  dimensions.
\newblock {\em Phys. Rev. D} {\bf 2012}, {\em 85},~104043.

\bibitem[Gunasekaran \em{et~al.}(2012)Gunasekaran, Kubiz{\v{n}}{\'a}k, and
  Mann]{gunasekaran2012extended}
Gunasekaran, S.; Kubiz{\v{n}}{\'a}k, D.; Mann, R.B.
\newblock Extended phase space thermodynamics for charged and rotating black
  holes and Born-Infeld vacuum polarization.
\newblock {\em J. High Energy Phys.} {\bf 2012}, {\em 2012},~1--43.

\bibitem[Allahverdizadeh \em{et~al.}(2013)Allahverdizadeh, Lemos, and
  Sheykhi]{allahverdizadeh2013extremal}
Allahverdizadeh, M.; Lemos, J.P.; Sheykhi, A.
\newblock Extremal Myers-Perry black holes coupled to Born-Infeld
  electrodynamics in five dimensions.
\newblock {\em Phys. Rev. D} {\bf 2013}, {\em 87},~084002.

\bibitem[Diaz-Alonso and Rubiera-Garcia(2013)]{diaz2013thermodynamic}
Diaz-Alonso, J.; Rubiera-Garcia, D.
\newblock Thermodynamic analysis of black hole solutions in gravitating
  nonlinear electrodynamics.
\newblock {\em Gen. Relativ. Gravit.} {\bf 2013}, {\em
  45},~1901--1950.

\bibitem[Ruffini \em{et~al.}(2013)Ruffini, Wu, and Xue]{ruffini2013einstein}
Ruffini, R.; Wu, Y.B.; Xue, S.S.
\newblock Einstein-Euler-Heisenberg theory and charged black holes.
\newblock {\em Phys.~Rev.~D} {\bf 2013}, {\em 88},~085004.

\bibitem[Zou \em{et~al.}(2014)Zou, Zhang, and Wang]{zou2014critical}
Zou, D.C.; Zhang, S.J.; Wang, B.
\newblock Critical behavior of Born-Infeld AdS black holes in the extended
  phase space thermodynamics.
\newblock {\em Phys. Rev. D} {\bf 2014}, {\em 89},~044002.

\bibitem[Cembranos \em{et~al.}(2015)Cembranos, de~la Cruz-Dombriz, and
  Jarillo]{cembranos2015reissner}
Cembranos, J.A.R.; De~la Cruz-Dombriz, A.; Jarillo, J.
\newblock Reissner-Nordstr{\"o}m black holes in the inverse electrodynamics
  model.
\newblock {\em J. Cosmol. Astropart. Phys.} {\bf 2015}, {\em
  2015},~042.

\bibitem[Jackson(1998)]{jackson1998classical}
Jackson, J.D.
\newblock {\em Classical Electrodynamics}, 3rd ed.; John Wiley: New York,  NY, USA, 1998.

\bibitem[Gibbons and Hawking(1977)]{gibbons1977cosmological}
Gibbons, G.W.; Hawking, S.W.
\newblock Cosmological event horizons, thermodynamics, and particle creation.
\newblock {\em Phys. Rev. D} {\bf 1977}, {\em 15},~2738--2751.

\bibitem[Hawking(1975)]{hawking1975particle}
Hawking, S.W.
\newblock Particle creation by black holes.
\newblock {\em Commun. Math. Phys.} {\bf 1975}, {\em
  43},~199--220.

\bibitem[Hartle and Hawking(1976)]{hartle1976path}
Hartle, J.B.; Hawking, S.W.
\newblock Path-integral derivation of black-hole radiance.
\newblock {\em Phys. Rev. D} {\bf 1976}, {\em 13},~2188--2203.

\bibitem[Gibbons and Hawking(1977)]{gibbons1977action}
Gibbons, G.W.; Hawking, S.W.
\newblock Action integrals and partition functions in quantum gravity.
\newblock {\em Phys.~Rev.~D} {\bf 1977}, {\em 15},~2752--2756.

\bibitem[Hawking(1978)]{hawking1978quantum}
Hawking, S.W.
\newblock Quantum gravity and path integrals.
\newblock {\em Phys. Rev. D} {\bf 1978}, {\em 18},~1747--1753.

\bibitem[Gibbons and Perry(1978)]{gibbons1978black}
Gibbons, G.; Perry, M.
\newblock Black holes and thermal Green functions.
\newblock  \emph{Proc. R. Soc. Lond. A  Math. Phys.
    Eng. Sci.}  \textbf{1978},  \emph{358}, 467--494.

\bibitem[Gibbons and Hawking(1993)]{gibbons1993euclidean}
Gibbons, G.W.; Hawking, S.W.
\newblock {\em Euclidean Quantum Gravity}; World Scientific: {Singapore, Singapore},~1993.

\bibitem[Caldarelli \em{et~al.}(2000)Caldarelli, Cognola, and
  Klemm]{caldarelli2000thermodynamics}
Caldarelli, M.M.; Cognola, G.; Klemm, D.
\newblock Thermodynamics of Kerr-Newman-AdS black holes and conformal field
  theories.
\newblock {\em Class. Quantum Gravity} {\bf 2000}, {\em 17},~399--420.

\bibitem[Bekenstein(1973)]{bekenstein1973black}
Bekenstein, J.D.
\newblock Black holes and entropy.
\newblock {\em Phys. Rev. D} {\bf 1973}, {\em 7},~2333--2346.

\bibitem[Wald(1993)]{wald1993black}
Wald, R.M.
\newblock Black hole entropy is the Noether charge.
\newblock {\em Phys. Rev. D} {\bf 1993}, {\em 48},~R3427--R3431.

\bibitem[Hawking and Page(1983)]{hawking1983thermodynamics}
Hawking, S.W.; Page, D.N.
\newblock Thermodynamics of black holes in anti-de Sitter space.
\newblock {\em Commun.~Math.~Phys.} {\bf 1983}, {\em
  87},~577--588.

\bibitem[Jardim \em{et~al.}(2012)Jardim, Rodrigues, and
  Houndjo]{jardim2012thermodynamics}
Jardim, D.F.; Rodrigues, M.E.; Houndjo, S.J.
\newblock Thermodynamics of phantom Reissner-Nordstrom-\linebreak AdS black hole.
\newblock {\em   Eur. Phys. J. Plus} {\bf 2012}, {\em
  127},~1--14.

\bibitem[Niu \em{et~al.}(2012)Niu, Tian, and Wu]{niu2012critical}
Niu, C.; Tian, Y.; Wu, X.N.
\newblock Critical phenomena and thermodynamic geometry of
  Reissner-\linebreak Nordstr{\"o}m-anti-de Sitter black holes.
\newblock {\em Phys. Rev. D} {\bf 2012}, {\em 85},~024017.

\bibitem[Cembranos \em{et~al.}(2014)Cembranos, De~La Cruz-Dombriz, and
  Jimeno~Romero]{cembranos2014kerr}
Cembranos, J.A.R.; De~la Cruz-Dombriz, A.; Jimeno~Romero, P.
\newblock Kerr-Newman black holes in f(R) theories.
\newblock {\em Int. J. Geom. Methods   Mod. Phys.}
  {\bf 2014}, {\em 11},~1450001.

\end{thebibliography}
\bibliographystyle{mdpi}

%%%%%%%%%%%%%%%%%%%%%%%%%%%%%%%%%%%%%%%%%%

%\abbreviations{Abbreviations/Nomenclature}
%
%Main text.

%%%%%%%%%%%%%%%%%%%%%%%%%%%%%%%%%%%%%%%%%%

%\appendix
%\section{Appendix Title}
%
%Main text.

\end{document}